\documentclass[11pt]{article}
\usepackage{amsmath,amssymb,color,epsfig,cite}
\usepackage{graphicx}
\usepackage{subfigure}
\usepackage{setspace}

\usepackage{amsfonts}

\newcommand{\hoch}[1]{$\, ^{#1}$}

\makeatletter
\@addtoreset{equation}{section}
\makeatother

\newcommand{\be}{\begin{equation}}
\newcommand{\ee}{\end{equation}}
\newcommand{\bea}{\setlength\arraycolsep{2pt} \begin{eqnarray}}
\newcommand{\eea}{\end{eqnarray}}
\newcommand{\nn}{\nonumber}

\def\ft#1#2{{\textstyle{\frac{\scriptstyle #1}{\scriptstyle #2} } }}
\def\fft#1#2{{\frac{#1}{#2}}}

\def\0{{\sst{(0)}}}
\def\1{{\sst{(1)}}}
\def\2{{\sst{(2)}}}
\def\3{{\sst{(3)}}}
\def\4{{\sst{(4)}}}
\def\5{{\sst{(5)}}}
\def\6{{\sst{(6)}}}
\def\7{{\sst{(7)}}}
\def\8{{\sst{(8)}}}
\def\9{{\sst{(9)}}}

\def\sst#1{{\scriptscriptstyle #1}}

\begin{document}



\begin{center}
{\large {\bf Scale Invariance vs. Conformal Invariance:\\ Holographic Two-Point Functions in Horndeski Gravity
}}

\vspace{10pt}
Yue-Zhou Li\hoch{1\dag},  H. L\"u\hoch{1*} and Hao-Yu Zhang\hoch{2\ddag}

\vspace{15pt}

\hoch{1}{\it Center for Joint Quantum Studies and Department of Physics,\\
School of Science, Tianjin University, Tianjin 300350, China}

\vspace{10pt}
\hoch{2}{\it Department of Physics, Shandong University, Jinan, Shandong 250100, China}

\vspace{30pt}

\underline{ABSTRACT}
\end{center}

We consider Einstein-Horndeski gravity with a negative bare constant as a holographic model to investigate whether a scale invariant quantum field theory can exist without the full conformal invariance. Einstein-Horndeski gravity can admit two different AdS vacua. One is conformal, and the holographic two-point functions of the boundary energy-momentum tensor are the same as the ones obtained in Einstein gravity. The other AdS vacuum, which arises at some critical point of the coupling constants, preserves the scale invariance but not the special conformal invariance due to the logarithmic radial dependence of the Horndeski scalar. In addition to the transverse and traceless graviton modes, the theory admits an additional trace/scalar mode in the scale invariant vacuum. We obtain the two-point functions of the corresponding boundary operators. We find that the trace/scalar mode gives rise to an non-vanishing two-point function, which distinguishes the scale invariant theory from the conformal theory. The two-point function vanishes in $d=2$, where the full conformal symmetry is restored. Our results indicate the strongly coupled scale invariant unitary quantum field theory may exist in $d\ge 3$ without the full conformal symmetry.  The operator that is dual to the bulk trace/scalar mode however violates the dominant energy condition.

\vfill {\footnotesize \hoch{\dag}liyuezhou@tju.edu.cn\ \ \ \hoch{*}mrhonglu@gmail.com\ \ \
\hoch{\ddag}zhanghaoyu11@icloud.com}

\pagebreak

\tableofcontents
\addtocontents{toc}{\protect\setcounter{tocdepth}{2}}


\newpage
\section{Introduction}
\label{sec:intro}

Conformal groups are generated by three types of transformations: (1) Poincar\'e transformations, (2) a scale (dilatation) transformation and (3) special conformal transformations. The Poincar\'e invariance is the underlying symmetry of any relativistic quantum field theories (QFT). Interestingly the Poincar\'e and scale transformations form a subgroup, which leads to an important question whether there exists a scale invariant quantum field theory (SQFT) that is not a (fully) conformal field theory (CFT).  After decades of research, a definite answer to this question for the general situation remains elusive. The subject has been reviewed in \cite{Nakayama:2013is} not so recently.  In $d=2$ dimensions, unitary scale invariant theories that have the discrete spectrum and the finite two-point function of energy-momentum tensor are necessarily conformal \cite{Polchinski:1987dy}\footnote{However, it was argued recently that from the mathematics viewpoint, the proof given by Polchinski \cite{Polchinski:1987dy} is incomplete, see \cite{Morinelli:2018pof}.}. Examples of SQFTs without a full conformal symmetry in $d=2$ violating these assumptions can be constructed, see, e.g. \cite{Iorio:1996ad,Riva:2005gd,Ho:2008nr}. In other dimensions, for example, in $d=3$ and $d\geq 5$, some so called ``free Maxwell theories'' were constructed and demonstrated that they were SQFTs but not CFTs \cite{ElShowk:2011gz}. However, the situation is much subtler in $d=4$. The perturbative approach can be used to demonstrate the enhancement of conformal symmetry from scale invariance near the fixed point\cite{Luty:2012ww}, and a number of such perturbative examples were studied extensively \cite{Fortin:2012cq,Fortin:2012hn,Fortin:2014tla}. It was argued that even beyond perturbative region, SQFT should also be CFT \cite{Luty:2012ww,Dymarsky:2013pqa,Naseh:2016maw}. However, no well-defined proof is available yet for the non-perturbative statement and a complete answer is far from clear in $d=4$.

The AdS/CFT correspondence \cite{Maldacena:1997re} provides a powerful tool to study certain strongly-coupled CFTs.  It is also natural to adopt the holographic technique for the SQFTs \cite{Nakayama:2010zz}. In fact the holographic approach may be exactly the right tool to address whether SQFTs without the full conformal invariance can exist, since this may be an intrinsic non-perturbative problem. Indeed, although anti-de Sitter (AdS) spacetimes with full conformal group arise naturally and commonly as vacua in bulk gravity theories, geometries that preserve both the Poincar\'e and scale invariance, but not full conformal invariance, are hard to come by.\footnote{Although Einstein gravity with minimally coupled vector fields can admit AdS that is not fully conformal, the theories typically violate the null-energy condition \cite{Nakayama:2010zz}.} The difficulty is a reflection of the fact that an SQFT is likely a CFT. However, concrete such an example does exist and it is provided by Einstein-Horndeski gravity coupled to a negative cosmological constant. Horndeski terms are higher-derivative invariant polynomials that are built from the Riemann curvature tensor and the 1-form of an axion \cite{Horndeski:1974wa,Nicolis:2008in}, analogous to the Gauss-Bonnet combination \cite{ll}.  It turns out that in addition to the usual AdS vacuum with the vanishing Horndeski scalar, the theory also admits the planar AdS at some critical point of the coupling constants, where the Horndeski scalar is non-vanishing and the special conformal invariance of the AdS is broken by the scalar. Black holes of Horndeski gravity at the critical point were also constructed, e.g. \cite{Rinaldi:2012vy,Anabalon:2013oea}.

The AdS/CFT and AdS/CMT applications of Einstein-Horndeski gravities have been extensively studied recently \cite{Feng:2015oea,Jiang:2017imk,Baggioli:2017ojd,Caceres:2017lbr,Liu:2017kml,Li:2018kqp,Liu:2018hzo,Feng:2018sqm}.  An important result is that the $a$-theorem can only be established for the scale invariant AdS vacuum, but not for the conformal AdS vacuum \cite{Li:2018kqp}.  This result strongly suggests that Einstein-Horndeski gravity on the critical AdS vacuum may provide a consistent holographic dual for some strongly coupled SQFT that is not conformal.  An important test to distinguish a CFT and SQFT is to examine the trace of the stress tensor, which vanishes for CFT, but not for SQFT.  The holographic dictionary \cite{Gubser:1998bc,Witten:1998qj} provides a powerful technique to calculate the two-point functions of the energy-momentum tensor of strongly-coupled field theory using classical gravity.  In this paper, we employ this technique to calculate the holographic two-point functions in Einstein-Horndeski gravity in both the conformal AdS and scale-invariance AdS vacua.  The holographic two-point functions in the conformal invariant vacuum satisfies
\be
\langle T_i^i(x) T_{jk} \rangle = 0 = \langle T_{jk}(x) T_i^i(0)\rangle\,,
\ee
as one would expect. However, the above quantities do not vanish for the scale invariant vacuum since there is in addition a trace/scalar mode.

The paper is organized as follows. In section \ref{Horndeski}, we review the Horndeski gravity and its vacuum solutions. One vacuum has full conformal symmetry, while the other at the critical point exhibits only the scale invariance. In section \ref{2-pt-conf}, we readily obtain the two-point functions of energy-momentum tensor in conformal vacuum, the result makes no difference compared to pure Einstein gravity. In section \ref{1-pt-trace}, we consider the scale invariant vacuum, and we obtain the linear perturbation solutions. We find in addition to the graviton modes, extra trace mode is also available. Furthermore, we obtain the holographic counterterms at the critical point. With the counterterms in hand, we employ the holographic dictionary to derive the two-point functions of the boundary energy-momentum tensor associated with the graviton modes and moreover, the one-point function formulae associated with the trace mode. In section \ref{2-pt-trace}, we analyze the ambiguity in determining the source of the trace mode. To resolve the ambiguity, we come up with an algebraic proposal to obtain the two-point functions. We also apply the extended metric basis method to verify the two-point functions we derive. In section \ref{d=2}, we discuss two-point functions in $d=2$, and it turns out the trace of energy-momentum tensor indeed gives no contribution to the two-point functions. The paper is summarized in section \ref{conclusion}. In Appendix \ref{graviton-ex}, we exhibits the algebraic proposal for the diagonal part of the graviton modes in Einstein gravity, and demonstrate explicitly that the algebraic proposal yields the right answer.

\section{Einstein-Horndeski gravity and AdS vacua}
\label{Horndeski}
In this paper, we consider Einstein gravity with a bare cosmological constant $\Lambda_0$, extended with the lowest-order Horndeski term.  The bulk action is given by
\bea
S_{\rm bulk} &=& \fft{1}{16\pi} \int d^D x \sqrt{-g}\, L_{\rm bulk}\,,\nn\\
L_{\rm bulk} &=& R - 2 \Lambda_0 - \ft12\alpha (\partial \chi)^2 + \ft12\gamma G_{\mu\nu} \partial^\mu \chi
\partial^\nu \chi\,.\label{genlag}
\eea
In this action, the Newton's constant is set to unity.  The remaining nontrivial parameter is the bare cosmological constant $\Lambda_0$ and the ratio $\gamma/\alpha$. This is because a constant scaling of the Horndeski scalar $\chi$ yields a homogeneous scaling of both the coupling constants $(\alpha,\gamma)$.  Note also that $\chi$ is axionic-like and the Lagrangian is invariant under the constant shift of $\chi$.

The covariant equations of motion of (\ref{genlag}) can be found in, e.g.~\cite{Anabalon:2013oea,Feng:2015oea}. They are given by the Einstein equation $E_{\mu\nu}=0$ and the scalar equation $E=0$, where \cite{Feng:2015oea}
\bea
E_{\mu\nu} &=& G_{\mu\nu} +\Lambda g_{\mu\nu} -
\ft12\alpha \Big(\partial_\mu \chi \partial_\nu \chi - \ft12 g_{\mu\nu} (\partial\chi)^2\Big)-\ft12\gamma \Big(\ft12\partial_\mu\chi \partial_\nu \chi R - 2\partial_\rho
\chi\, \partial_{(\mu}\chi\, R_{\nu)}{}^\rho \cr
&&- \partial_\rho\chi\partial_\sigma\chi\, R_{\mu}{}^\rho{}_\nu{}^\sigma -
(\nabla_\mu\nabla^\rho\chi)(\nabla_\nu\nabla_\rho\chi)+(\nabla_\mu\nabla_\nu\chi)
\Box\chi + \ft12 G_{\mu\nu} (\partial\chi)^2\cr
&&-g_{\mu\nu}\big[-\ft12(\nabla^\rho\nabla^\sigma\chi)
(\nabla_\rho\nabla_\sigma\chi) + \ft12(\Box\chi)^2 -
  \partial_\rho\chi\partial_\sigma\chi\,R^{\rho\sigma}\big]\Big)\,,\cr
E &=&\nabla_\mu \big( (\alpha g^{\mu\nu} - \gamma G^{\mu\nu}) \nabla_\nu\chi\big)\,.
\eea
It is easy to see that the theory admits the AdS vacuum of radius $\ell$, namely
\be
ds_D^2 = ds_{\rm AdS_{d+1}}^2\,,\qquad \chi=0\,;\qquad \Lambda_0 =-\fft{d(d-1)}{2\ell^2}\,,\qquad D=d+1\,.
\label{genericads}
\ee
This vacuum involves only the Einstein gravity sector, and the Horndeski term can be treated perturbatively for small $\gamma$. The linearized perturbation of the scalar $\chi$ has the kinetic term
\be
-\ft12 \Big(\alpha -\ft12 d(d-1)\gamma \ell^{-2}  \Big) (\partial\chi)^2\,.
\ee
The ghost-free condition requires that
\be
\alpha - \ft12 d (d-1)\gamma\ell^{-2} \ge 0\,.
\ee
At the saturation point of the above inequality, which is referred to as the ``critical point'' in \cite{Li:2018kqp}, the theory admits a new AdS vacuum, whose radius is not governed by the bare cosmological constant, but by the ratio $\gamma/\alpha$ instead:
\bea
ds_D^2 &=& \fft{\ell^2 dr^2}{r^2} + r^2 \eta_{ij} dx^i dx^j\,,\qquad
 \chi=\chi_s \log(\fft{1}{r^2}) + \chi_0\,,\nn\\
\fft{\alpha}{\gamma}&=&\fft{d(d-1)}{2\ell^2}\,,\qquad
\Lambda_0 =-\fft{d(d-1)}{2\ell^2}\Big(1 + \fft{2\gamma \chi_s^2}{\ell^2}\Big)\,.
\label{criticalads}
\eea
Note that the $\chi$ solution is parameterized to be the same as in \cite{Li:2018kqp}.  In this vacuum solution, the Horndeski term is an integral part and cannot be viewed as a small perturbation.  In particular, when $\ell$ is large, corresponding to small curvature in gravity, we must have large $\gamma$; on the other hand, the small $\gamma$ implies small $\ell$ and the corresponding large spacetime curvature.  A further important feature is that the full AdS conformal symmetry of the metric is broken by the Horndeski axion down to the subgroup of Poincar\'e symmetry together with the scale invariance, namely
\footnote{We do do not consider in this paper the scale transformation that leaves the equations of motion invariant, but not the action. Such scale invariance was referred to as the ``trombone'' symmetry in\cite{Cremmer:1997xj}.}
 \be
x^i\rightarrow \lambda x^i\,,\qquad r^2\rightarrow \fft{r^2}{\lambda}\,.
\ee
The special conformal transformation invariance of the conformal group is broken.
It should be pointed out right away that under the above scale transformation, the axion $\chi$ undergoes a constant shift. We take the view that since $\chi$ appears in the theory only through a derivative, vacua with $\chi$ and $\chi+c$ for any constant $c$ should be identified as the same. The holographic dual is thus expected to be a relativistic quantum field theory with the scale rather than the full conformal invariance.

Holographic conformal anomaly for Einstein-Horndeski gravity (\ref{genlag}) was recently obtained \cite{Li:2018kqp}. Specifically, the $a$-charges are
\bea
\hbox{generic AdS vacuum}:&&\qquad a= \ell^{d-1}\,,\nn\\
\hbox{critical AdS vacuum}:&&\qquad a= \ell^{d-1} \Big(1 - (d-1)\fft{\gamma \chi_s^2}{\ell^2}\Big)\,.\label{central}
\eea
Note that we have stripped off the overall (inessential) purely numerical constants in presenting the $a$-charges. It turns out that the $a$-theorem cannot be established for the generic AdS vacuum, but it can be for the critical vacuum \cite{Li:2018kqp}.

\section{Two-point functions in the conformal vacuum}
\label{2-pt-conf}
The linear perturbation of the generic AdS vacuum (\ref{genericads}) involves the graviton and axion modes.  Taking the AdS spacetime to be the planar type, the perturbations are
\be
ds^2 = \fft{\ell^2 dr^2}{r^2} + r^2 \eta_{ij} dx^i dx^j + h_{ij} dx^i dx^j\,,\qquad
\chi=\chi(r,x)\,.
\ee
This is the linearized version of the Fefferman-Graham (FG) expansion with the gauge choice $h_{\mu i}=0$, where $\mu=(r,i)$. The physical spin-2 massless graviton mode satisfies the further transverse and traceless conditions, which imply
\be
\partial^i h_{ij} =0\,,\qquad h_{ij} \eta^{ij}=0\,,\label{tt}
\ee
where $\partial^i=\eta^{ij} \partial_j = \eta^{ij} \partial/\partial x^j$.  The perturbation satisfies
\be
(\bar\Box + \fft{2}{\ell^2} ) h_{ij}=0\,,
\ee
where the Laplacian $\bar\Box$ is defined with respect to the AdS vacuum. The scalar field satisfies
\be
\Big(\alpha -\ft12 d(d-1)\gamma \ell^{-2}  \Big)\,\bar \Box \chi=0\,.
\ee
Thus the linear equation of the graviton is identical to that of AdS vacuum in Einstein gravity, and the solution can be expressed in terms of the Hankel's function
\be
h_{ij} = c_{ij} r^{-\fft{d-4}{2}} H_{\fft{d}2} (\fft{p\,\ell}{r})\, e^{{\rm i} p\cdot x}\,,\qquad p^2=-p^i p^j \eta_{ij}\,,\qquad
c_{i t}=0\,,\qquad c_{ij} \eta^{ij}=0\,.
\ee
This leads to the holographic two-point function for the boundary energy-momentum tensor \cite{Liu:1998bu}
\be
\langle T_{ij}(x)T_{kl}(0)\rangle =\fft{C_T\,
\mathcal{I}_{ijkl}(x)}{x^{2d}}\,,
\label{conformal2pt}
\ee
where $\mathcal{I}_{ijkl}(x)$ is the boundary spacetime tensor defined by
\bea
\mathcal{I}_{ijkl}(x)=\ft{1}{2}\big(I_{ik}(x)I_{jl}(x)+I_{il}(x)I_{jk}(x)\big)-
\ft{1}{d}\eta_{ij}\eta_{kl}\,,\qquad I_{ij}(x)=\eta_{ij}-\fft{2x_{i}x_{j}}{x^2}\,.\label{Iijkl}
\eea
The coefficient $C_T$ can be expressed as \cite{Li:2018drw}
\be
C_T = N_2 {\cal C}_T\,,\qquad \hbox{with}\qquad {\cal C}_T = \ell^{d-1}\,,\qquad N_2 = \fft{\Gamma(d+2)}{8 (-1)^{\fft{d}2}\pi^{\fft{d}2 +1} (d-1)\Gamma(\fft{d}2)}\,.
\ee
In other words, ${\cal C}_T$ is the dimensionful quantity with the overall purely numerical factors stripped off.
It is easy to see the identity
\be
{\cal C}_T = \fft{1}{d-1} \ell \fft{\partial a}{\partial \ell}\,.
\ee
This identity was seen and conjectured to hold for all the higher-order massless gravities \cite{Li:2018drw}.

\section{Linear modes and boundary terms in the scale invariant vacuum}
\label{1-pt-trace}
We now consider the linear perturbation of the AdS vacuum at the critical point (\ref{criticalads}).  The ansatz is again the linearized FG type, namely
\be
ds^2 = \fft{\ell^2 dr^2}{r^2} + r^2 \eta_{ij} dx^i dx^j + h_{ij} dx^i dx^j \,,\qquad
\chi=\chi_s \log\big(\fft{1}{r^2}\big) + \psi(r,x)\,.
\ee
The linearized equations are complicated and they become simpler when we examining the transverse traceless modes and the trace mode separately.

\subsection{Graviton mode}
\label{Gravi-linear}

The massless spin-2 graviton mode satisfies further the transverse and traceless conditions.  The scalar equation is then automatically satisfied at the linear level.  The tensor $E_{\mu\nu}$ becomes
\bea
E_{\mu\nu}&=&-\ft{1}{2}(1-\frac{\gamma \chi_s^2}{\ell^2})(\Box + \fft{2}{\ell^2}) h_{\mu\nu}-\frac{d \gamma \chi_s^2}{\ell^3} n^\rho \nabla_\rho h_{\mu\nu}-\frac{ \gamma \chi_s^2}{\ell^2} n^\rho n^\sigma \nabla_\rho\nabla_\sigma h_{\mu\nu}\cr
&&+\frac{2(d-1)\chi_s\gamma}{\ell^3} n_{(\mu} \nabla_{\nu)} \psi-\frac{2\chi_s\gamma}{\ell^2}\beta_{\rho(\mu}\nabla_{\nu)}\nabla^\rho\psi+\frac{\gamma d \chi_s}{\ell^2}\nabla_\mu\nabla_\nu \psi+\frac{\gamma \chi_s}{\ell^2}\beta_{\mu\nu}\Box \psi\cr
&&-g_{\mu\nu}\Big(\fft{2 (d-1)\gamma \chi_s}{\ell^3}n^\rho\nabla_\rho \psi -\frac{\chi_s\gamma}{\ell^2}\beta_{\rho\sigma}\nabla^\rho\nabla^\sigma\psi+\frac{d\gamma}{\ell^2}\Box \psi\Big)\,,
\eea
where the metric and covariant derivatives are defined on the AdS background, and
\be
\beta_{\mu\nu} \equiv g_{\mu\nu} - n_\mu n_\nu=\ell \nabla_{(\mu} n_{\nu)}\,,\qquad n\equiv n^\mu \partial_\mu = \fft{r}{\ell} \fft{\partial}{\partial r}\,.
\ee
The linearized equation $E_{r i}=0$ implies that
\be
\fft{(d-1) \gamma \chi_s }{\ell^2} \partial_i \partial_r\psi=0\,.
\ee
Thus we have $\psi = c_r r + c_i x^i$.  The equation $E_{rr}=0$ implies that
\be
-\fft{(d-1)\gamma \chi_s}{\ell^2 r} \big(d\partial_r \psi + \fft{\ell^2}{r^3} \eta^{ij} \partial_i \partial_j\psi\big)=0\,.
\ee
Thus $c_r=0$ and $\psi=c_i x^i$.  It follows that the equation $E_{ij}=0$ does not involve $\psi$ and we have
\be
(1 + \fft{\gamma \chi_s^2}{\ell^2}) \big(\bar \Box + \fft{2}{\ell^2}\big) h_{ij} -\fft{2\gamma \chi_s^2}{\ell^2 r^2} \Box h_{ij}=0\,,
\ee
where $\Box$ is defined with respect to the metric $\eta_{ij}$.  The absence of ghost excitations requires that the coefficient of the time-derivative term $\ddot h_{ij}$ be non-negative, namely
\be
\kappa_{\rm eff} = 1 - \fft{\gamma \chi_s^2}{\ell^2}\ge 0\,.\label{kapeff}
\ee
The solution is given by
\be
h_{ij} = c_{ij} r^{-\fft{d-4}{2}} H_{\fft{d}{2}}\big(\fft{b \ell p}{r}\big)\, e^{{\rm i} p\cdot x}\,,\qquad
c_{it}=0\,,\qquad c_{ij} \eta^{ij}=0\,,\qquad b= \sqrt{\fft{\ell^2 -\gamma \chi_s^2}{\ell^2 + \gamma \chi_s^2}}\,.
\ee

\subsection{The trace mode}
\label{Trace-linear}

In addition to the transverse and traceless graviton mode obtained in the previous subsection, we find the theory admits an additional scalar mode that consists of the metric trace and also the Horndeski axion excitation.  Taking the Lorenz gauge $p^\mu = (-E, 0,0,\ldots,0)$, the ansatz is given by
\be
ds^2 =\fft{\ell^2 dr^2}{r^2} + r^2 \eta_{ij} dx^i dx^j + h_0 dt^2 +
h \delta_{\tilde i \tilde j}dx^{\tilde i} dx^{\tilde j}\,,\qquad \chi = \chi_s \log (\fft{1}{r^2}) + \psi
\ee
The perturbative functions $(h_0,h,\psi)$ depend on the bulk radius $r$ and boundary time $t$ only. Thus any constant $r$ slice of the spacetime is an FLRW cosmological metric.  For this reason we may also call this the cosmological mode.  The kinetic term of the linearized bulk Lagrangian is
\bea
L&=&\fft{(d-1)(d-2)}{64\pi}\ell r^{d-7} \Big(-\ell^2 \kappa_{\rm eff} \dot {\tilde h}^2 +
\fft{16\gamma^2 \chi_s^2 r^4}{(d-2)^2 \ell^2 \kappa_{\rm eff}} \dot\psi^2\Big)\,,\nn\\
\tilde h &=& h - \fft{4\gamma  \chi_s r^2}{(d-2) \ell^2\kappa_{\rm eff}^2} \psi\,,
\eea
where $\kappa_{\rm eff}$ is given in (\ref{kapeff}). This Lagrangian is analogous to the kinetic term of the linearized FLRW model with $\tilde h$ corresponding to the scale factor and $\psi$ corresponding to the matter scalar field.  The absence of the ghost excitation requires that $\kappa_{\rm eff}\ge 0$.

The full linearized equations can be solved by
\be
h_0 = -\frac{2 \gamma \chi _s \left(\ell^2-(2d-3) \gamma  \chi _s^2\right)}{\left(\ell^2+\gamma  \chi _s^2\right) \left(\ell^2+ 3 \gamma  \chi _s^2\right)} r^2\,\psi\,,\qquad
h = \frac{2 \gamma \chi _s}{\ell^2+\gamma  \chi _s^2} r^2\,\psi\,,\label{h0h}
\ee
where $\psi$ satisfies
\be
d(\ell^2 + \gamma \chi_s^2) r^3 (r\psi'' + (d+1) \psi') - (d-2)\ell^2(\ell^2 + 3 \gamma \chi_s^2)\ddot\psi=0
\ee
The general solution is given by
\be
\psi=r^{-\fft{d}2} H_{\fft{d}2}(\fft{\tilde b \ell E}{r})\, e^{-{\rm i} E t}\,,\qquad
\tilde b=\sqrt{\ft{(d-2)(\ell^2+3\gamma \chi_s^2)}{d\,(\ell^2 + \gamma \chi_s^2)}}\,.\label{psi}
\ee
Performing a Lorentz transformation in the boundary, we obtain the Lorentz covariant expression for the trace/scalar mode, namely
\be
ds^2 =\fft{\ell^2 dr^2}{r^2} + r^2 \eta_{ij} dx^i dx^j + h_{ij} dx^i  dx^j\,,\qquad \chi = \chi_s \log (\fft{1}{r^2}) + \psi\,,
\ee
where the Lorentz covariant linear perturbation is
\be
h_{ij} = -\Big(\fft{p_i p_j}{p^2} - \eta_{ij}\Big) h - \fft{p_i p_j}{p^2} h_0\,,\label{tracehij}
\ee
and $(h_0,h)$ are given by (\ref{h0h}) with (\ref{psi}) where $Et$ is replaced by $p_i x^i$.  It is easy to verify that this mode is neither transverse nor traceless; it is the Lorentz covariantization of the boundary cosmological mode.

\subsection{Boundary action}

In order to derive the boundary properties from the bulk perturbations, it is necessary to construct the boundary action.  The boundary action contains two parts.  The first part is the Gibbons-Hawking surface term, which is given by
\be
S_{\rm surf} = \fft{1}{8\pi} \int d^d x \sqrt{-h} \Big[K + \ft14\gamma \big(
\partial_\mu \chi\partial_\nu \chi\, n^\mu n^\nu - (\partial\chi)^2\big)\,K + \ft14\gamma
\partial_\mu \chi \partial_\nu \chi K^{\mu\nu}\Big]\,,
\ee
where $n^\mu$ is the unit vector normal to the surface and $K$ is the trace of the second fundamental form $K_{\mu\nu} = h_{\mu}^\rho \nabla_\rho n_\nu$ and $h_{\mu\nu} = g_{\mu\nu} - n_\mu n_\nu$. The first term in the square bracket is the contribution from the Einstein-Hilbert term \cite{Gibbons:1976ue}. The $\gamma$-dependent terms are associated with the Horndeski term in the bulk action.  Note that in this subsection only, we use the standard convention $h_{\mu\nu}$ for the boundary metric. It was referred to as $\beta_{\mu\nu}$ in subsection \ref{Gravi-linear}. There should be no confusion between the $h_{\mu\nu}$ here and the metric perturbation $h_{ij}$ in the rest of the paper.

The second part is the boundary counterterm that is necessary for asymptotic AdS spacetimes. We find, up to the quadratic curvature term and scalar term, that it is given by
\be
S_{\rm ct} = \fft{1}{16\pi} \int d^dx \sqrt{-h} \big(c_0+ c_1 {\cal R} + c_2 {\cal R}^{ij} {\cal R}_{ij}
+c_3 {\cal R}^2 + b_1 (\partial_i\chi\partial^i\chi)+b_2 \partial_i\Box\chi\partial^i\chi+ \cdots\big)\,,
\ee
where ${\cal R}$ is the Ricci scalar of the boundary metric, and the derivatives are associated with the boundary metric. The coefficients of the counterterms for the generic AdS vacuum was know, given by
\bea
 c_0 &=& -\fft{2(d-1)}{\ell} \,,\quad c_1=-\fft{\ell}{d-2}\,,\quad c_2 = -\fft{\ell^3}{(d-4)(d-2)^2}\,,\quad
c_3=-\fft{d}{4(d-1)} c_2\,,\nn\\
\cr  b_1&=&b_2=0\,.
\eea
For the critical AdS vacuum, we find that they become
\bea
c_0 &=&-\fft{2(d-1)}{\ell} (1 + \fft{\gamma \chi_s^2}{\ell^2}) \,,\qquad
c_2 = -\fft{\ell^3}{(d-4)(d-2)^2}\fft{(1-\fft{\gamma \chi_s^2}{\ell^2})^2}{1 + \fft{\gamma \chi_s^2}{\ell^2}}\,,\nn\\
c_3 &=& -\fft{d}{4(d-1)} \Big(1 -\fft{2(d-2)^2 \gamma \chi_s^2 (\ell^2 + 3 \gamma \chi_s^2)}{
d^2 (\ell^2 - \gamma \chi_s^2)} \Big)c_2\,,\qquad c_1=-\fft{\ell}{d-2}(1 - \fft{\gamma \chi_s^2}{\ell^2})\,,\nn\\
\cr
 \quad b_1 &=& \fft{(d-1)\gamma}{2\ell}\,,\qquad
b_2= \fft{d-1}{d(d-4)}\fft{\ell\gamma^2\chi_s^2(\ell^2+3\gamma\chi_s^2)}{(\ell^2+\gamma\chi_s^2)^2}\,.
\eea

\subsection{Holographic one-point functions}
\label{holographic1pt}

Having obtained the full action, namely
\be
S_{\rm tot}=S_{\rm bulk} + S_{\rm surf} + S_{\rm ct}\,,
\ee
we are in the position to compute the holographic one-functions associated with the linear modes we obtained in section \ref{Gravi-linear} and \ref{Trace-linear}. From the Brown-York stress tensor associated with the perturbation mode $h_{ij}$,
\be
T_{ij}(h)=-\fft{2}{r^{d}}\fft{\delta S_{\rm tot}}{\delta h^{ij}}\,,\label{BY}
\ee
we can define the one-point function of the holographic energy-momentum tensor \cite{deHaro:2000vlm}
\be
\langle T_{ij}\rangle=T_{ij}(h)r^{d-2}\big|_{r\rightarrow\infty}\,.\label{1-pt-0}
\ee
The two-point function can be obtained from dividing the one-point function from the leading expression of the mode $h_{ij}$. This approach is not yet covariant and a full covariant approach based on writing the modes in the metric basis was given in \cite{Li:2018drw,Johansson:2012fs,Ghodsi:2014hua}.  In this paper, we shall not go through this detail but simply present the results.  The two-point function of the boundary energy-momentum tensor associated with the bulk graviton is given immediately by the same expression (\ref{conformal2pt}), with $C_T=N_2 {\cal C}_T$ and
\be
{\cal C}_T = \Big(\fft{\ell^2-\gamma \chi_s^2}{\ell^2 + \gamma \chi_s^2}\Big)^{\fft{d}{2}}
\Big(1 + \fft{\gamma\chi_s^2}{\ell^2}\Big)\ell^{d-1}\,.\label{CT-gravi}
\ee
Intriguingly, comparing to (\ref{central}), we have ${\cal C}_T =a+ (d-2) {\cal O}(\gamma^2)$. It is worth pointing out that as the dimension $d$ increases, more and more higher-order counterterms are necessary to cancel the divergence as $r\rightarrow \infty$.  However, only the coefficient $c_0$ contributes to the overall coefficient of the two-point functions.

To obtain the one-point functions associated with trace mode, we denote the corresponding source as $h_0^{(0)}$, $h^{(0)}$  and $\psi^{(0)}$ which are the corresponding leading terms in the expansion, i.e.
\bea
&& h_0=r^2(h_0^{(0)}+\cdots+\fft{h_0^{(d)}}{r^d}+\cdots)\,,\qquad h=r^2(h^{(0)}+\cdots+\fft{h^{(d)}}{r^d}+\cdots)\,,
\cr && \psi=\psi^{(0)}+\cdots+\fft{\psi^{(d)}}{r^d}+\cdots\,.\label{expansion}
\eea
By applying (\ref{BY}) and (\ref{1-pt-0}),we find that the one-point function of the energy-momentum tensor takes the form
\be
\langle T_{tt} \rangle =0\,,\qquad \langle T_{\tilde i \tilde j}\rangle =\delta_{\tilde i\tilde j}
\langle T \rangle\,,\qquad \langle T\rangle=\fft{d^2\gamma\chi_s^2}{8\pi\ell^3}\fft{\ell^2+\gamma\chi_s^2}{\ell^2
+3\gamma\chi_s^2}h^{(d)}\,.\label{1-pt-T}
\ee
The vanishing of $T_{tt}$ implies that the trace/scalar mode violates the dominant energy condition, but it can satisfy the other energy conditions. The result can be straightforwardly covariantized and then we have
\be
\langle T_{ij}\rangle=-\Big(\fft{p_i p_j}{p^2} - \eta_{ij}\Big) \langle T\rangle\,.\label{1-pt-T-cov}
\ee
The one-point function of $\chi$ can be defined similarly
\be
\langle\chi\rangle=\fft{\delta S}{\delta\psi^{(0)}}\,.
\ee
For now, we have
\be
\langle\chi\rangle=-\fft{\,d^2(d-1)\gamma^2\chi_s^2}{8\pi\ell^3(\ell^2+3\gamma\chi_s^2)}\label{1-pt-chi}
\psi^{(d)}\,.
\ee
It is of importance to keep in mind that from (\ref{h0h}) and (\ref{expansion}) we have
\be
h^{(0)} = \frac{2 \gamma \chi _s}{\ell^2+\gamma  \chi _s^2}\,\psi^{(0)}\,,\qquad
h^{(d)} = \frac{2 \gamma \chi _s}{\ell^2+\gamma  \chi _s^2}\,\psi^{(d)}\,.
\ee
It follows that we have
\be
\langle T\rangle=-\fft{2\chi_s}{d-1}\langle\chi\rangle\,,\label{T-chi-rela}
\ee
and they are linear functions of the source $h^{(0)}$ (or $\psi^{(0)}$.) It follows from (\ref{1-pt-T-cov}) that the trace of the energy-momentum tensor is
\be
\langle T_i{}^i\rangle = -2\chi_s\langle\chi\rangle\,.\label{trace}
\ee
Thus we see that the boundary scalar operator has the same scaling dimension as the stress tensor, namely $\Delta=d$. It should be commented that in QFT the non-vanishing of the stress tensor trace is by itself not necessary a definite indication of violation of the conformal symmetry. In fact, the energy-momentum tensor constructed in QFT is not unique and it can be redefined by introducing local counterterms in the Lagrangian. Given this crucial property, it turns out if one has $\langle T_i{}^i\rangle\propto \partial_i \partial_j L^{ij}$ for some local operator $L^{ij}$ of scaling dimension $\Delta=d-2$, one can introduce appropriate local counterterms in the Lagrangian to give a redefined energy-momentum tensor that is trace free.  However, it shall be emphasized that the situation is very different in the holographic context. The holographic dictionary provides an exclusive instruction for us to draw conclusions about the dual QFT by studying classical gravity. From the bulk point of view, the counterterms added in the boundary action in this section play the exclusive role of canceling the divergence, but they cannot contribute any further parts to modify the holographic energy-momentum tensor. Therefore, our one-point function of the holographic energy-momentum tensor has no ambiguity, and the result (\ref{trace}) reflects convincingly the breaking of the full conformal symmetry. Furthermore, it is clear that $\chi$ is a {\it fundamental} operator in the boundary field theory and it cannot be expressed in terms of derivatives of other operators, thus the conformal symmetry is broken by this scalar operator.  However, the expression (\ref{trace}) also shows that there is no local expression for a virial current, which was typically introduced in a scale invariant theory to form a conserved current associated with the scaling symmetry.  This seems to suggest that even the scaling symmetry may be broken.  It is thus necessary to study the two-point functions associated with the trace scalar mode and we find that the scaling symmetry does preserve. In the next subsection, we carrying out the computation of the two-point functions, including $\langle TT\rangle$, $\langle T\chi\rangle=\langle\chi T\rangle$ and $\langle\chi\chi\rangle$.

\section{Two-point functions in scale invariant vacuum}
\label{2-pt-trace}

The two-point functions of energy-momentum tensor associated with graviton modes can be obtained readily and it is given in (\ref{conformal2pt}) with the coefficient $C_{\rm T}$ given in (\ref{CT-gravi}). In this section, we focus on the two-point functions associated with the trace mode. However, as the on-shell solution (\ref{h0h}) suggests, all seemingly different sources, $h^{(0)}_{ij}$ and $\psi^{(0)}$ actually belong to the same singlet, so do the responses $h^{(d)}_{ij}$ and $\psi^{(d)}$. In fact they are the Lorentz covariantization of the cosmological mode. It is thus not apparent at the first sight which source contributes the specific percentage of the response in the one-point functions (\ref{1-pt-T}), (\ref{1-pt-T-cov}) and (\ref{1-pt-chi}), which is a necessary information to derive the corresponding two-point functions. The analogous stituation can be found even in graviton modes, which is discussed in Appendix \ref{graviton-ex}. Following the procedure presented in Appendix \ref{graviton-ex} for an simper example, we shall rewrite the one-point functions such that the distinctions between different contributions become clear.

\subsection{An algebraic proposal}
\label{subsec:algebraic}

It follows from (\ref{tracehij}) that the covariant source $h^{(0)kl}$ is given by
\be
h^{(0)}_{ij}=-\Big(\fft{p_i p_j}{p^2} - \eta_{ij}\Big) h^{(0)} - \fft{p_i p_j}{p^2} h_0^{(0)}\,.
\ee
Denoting
\be
\Theta_{ij}=p_i p_j-\eta_{ij}p^2\,,
\ee
we can rewrite the one-point functions of the previous subsection as
\be
\langle T_{ij}\rangle=\fft{a_1}{p^4} \Theta_{ij}\Theta_{kl}h^{(0)kl}+\fft{a_2}{p^2}\Theta_{ij}\psi^{(0)}\,,
\qquad \langle\chi\rangle=\fft{b_1}{2p^2}\Theta_{ij}h^{(0)ij}+b_2\psi^{(0)}\,.\label{1-pt-new}
\ee
In other words, we split the one-point functions as the linear combinations of the metric and scalar sources as if they have different origins, with the coefficients $a_i$, $b_i$ to be determined. Requiring $\langle T\chi\rangle=\langle \chi T\rangle$ implies $a_2=b_1$. The two-point functions can then be deduced, yielding
\bea
 \langle T_{ij}T_{kl}\rangle=\fft{2a_1}{p^4} \Theta_{ij}\Theta_{kl}\,,\qquad \langle\chi\chi\rangle=b_2\,,
\qquad \langle T_{ij}\chi\rangle=\langle\chi T_{ij}\rangle=\fft{a_2}{p^2}\Theta_{ij}\,.\label{2-pt-new}
\eea
To obtain $a_i$ and $b_i$, we note they must be such that (\ref{1-pt-new}) reproduces the results (\ref{1-pt-T}), (\ref{1-pt-T-cov}) and (\ref{1-pt-chi}). It follows from the identity
\be
\Theta_{ij}h^{(0)ij}=-(d-1)h^{(0)}p^2\,,\label{Theh}
\ee
that we have
\be
(d-1)a_1-\fft{\ell^2+\gamma\chi_s^2}{2\gamma\chi_s}a_2=\fft{\langle T\rangle}{h^{(0)}}\,,
\qquad -\fft{(d-1)\gamma\chi_s}{\ell^2+\gamma\chi_s^2}a_2+b_2=\fft{\langle\chi\rangle}{\psi^{(0)}}\,,\label{const-1}
\ee
where $\langle T\rangle$ and $\langle\chi\rangle$ should be evaluated using (\ref{1-pt-T}) and (\ref{1-pt-chi}). Furthermore, (\ref{2-pt-new}) forms a $2\times2$ matrix
\be
\left(
\begin{array}{cc}
 2a_1 & a_2 \\
 a_2 & b_2 \\
\end{array}
\right)\nn
\ee
that must have one zero eigenvalue since the system has only one independent mode. Hence, we have the constraint
\be
2a_1 b_2-a_2^2=0\,.\label{const-2}
\ee
Thus we have three equations for  $(a_1,a_2,b_2)$. Solving (\ref{const-1}) and (\ref{const-2}), we can obtain $a_i$ and $b_i$. They are given by
\bea
 a_1=\fft{2\gamma\chi_s^2}{(d-1)(\gamma\chi_s^2-\ell^2)}\fft{\langle T\rangle}{h^{(0)}}\,,\qquad a_2=b_1=
\fft{2\gamma\chi_s}{\gamma\chi_s^2-\ell^2}\fft{\langle T\rangle}{h^{(0)}}\,,
\qquad b_2=\fft{\ell^2+\gamma\chi_s^2}{\ell^2-\gamma\chi_s^2}\fft{\langle\chi\rangle}{\psi^{(0)}}\,.\label{ab-solu}
\eea
The zero eigenvector is given by
\be
\langle N_{ij}\rangle=\langle T_{ij} \rangle- \fft{2\chi_s}{(d-1)} \fft{\Theta_{ij}}{p^2} \langle \chi\rangle\,.\label{Nij}
\ee
Thus the operator $N_{ij}$ is null with $\langle N_{ij}\cdots\rangle =0$, and decouples from the physical spectrum, implying that trace/scalar operators have only one nontrivial combination. The result is consistent with (\ref{T-chi-rela}), reflecting that this is the right approach. In Appendix \ref{graviton-ex}, we use the same procedure to obtain the correct two-point functions of the diagonal part of the energy momentum tensor associated with the graviton mode. In the next subsection we verify the result by means of the similar analysis developped in \cite{Li:2018drw,Johansson:2012fs,Ghodsi:2014hua} and encoding the trace/scalar mode in the extended metric basis.

\subsection{Extended metric basis}
In this subsection, we aim to validate the algebraic proposal in the previous subsection by following and generalizing the explicit analysis in \cite{Li:2018drw,Johansson:2012fs,Ghodsi:2014hua}. Considering we have only one single mode, $h^{(0)}_{ij}$ and $\psi^{(0)}$ should no longer be separated, it is natural to define an extended metric to combine $h^{(0)}_{ij}$ and $\psi^{(0)}$ together, i.e.
\be
h^{(0)}_{ab}=(h^{(0)}_{ij},2\psi^{(0)})=h^{(0)}_{ij}e^{(i)}_a e^{(j)}_b+2\psi^{(0)}e^{(d)}_a e^{(d)}_b\,,
\ee
where we have introduced the extended vielbein $e^{(b)}_a$ with $a=0,\ldots, d$, and the factor ``$2$'' is for latter convenience. We can also define the extended one-point function
\be
\langle T_{ab}\rangle=(-\fft{\Theta_{ij}}{p^2}\langle T\rangle,\langle\chi\rangle)=-\fft{\Theta_{ij}}{p^2}\langle T\rangle e^{(i)}_a e^{(j)}_b+\langle\chi\rangle e^{(d)}_a e^{(d)}_b\,.
\ee
With this notation, the two-point functions are given by
\be
\langle T_{ab}T_{ef}\rangle=\fft{2\delta\langle T_{ab}\rangle}{\delta h^{(0)ef}}=
\left(
\begin{array}{cc}
 \langle T_{ij}T_{kl}\rangle &\,\, \langle T_{ij}\chi \rangle \\
 \langle \chi T_{ij}\rangle &\,\, \langle\chi\chi\rangle \ \\
\end{array}
\right)
\,.
\ee
To proceed, after using (\ref{Theh}), we note
\be
T_{ab}h^{(0)ab}=(d-1)\fft{\gamma\chi_s^2-\ell^2}{2\gamma\chi_s^2}\langle T\rangle h^{(0)}=
\fft{2(\ell^2-\gamma\chi_s^2)}{\ell^2+\gamma\chi_s^2}\langle\chi\rangle \psi^{(0)}:=C\,.\label{C}
\ee
Therefore, we have
\be
\langle T_{ab}\rangle=\Big(-\fft{\Theta_{ij}}{p^2}\langle T\rangle e^{(i)}_a e^{(j)}_b+\langle\chi\rangle e^{(d)}_a e^{(d)}_b\Big)\Big(-\fft{\Theta_{kl}}{p^2}\langle T\rangle e^{(k)}_e e^{(l)}_f+\langle\chi\rangle e^{(d)}_e e^{(d)}_f\Big)
\fft{h^{(0)ef}}{C}\,.
\ee
We end up with
\be
\langle T_{ab}T_{ef}\rangle=\fft{2}{C}\Big(-\fft{\Theta_{ij}}{p^2}\langle T\rangle e^{(i)}_a e^{(j)}_b+\langle\chi\rangle e^{(d)}_a e^{(d)}_b\Big)\Big(-\fft{\Theta_{kl}}{p^2}\langle T\rangle e^{(k)}_e e^{(l)}_f+\langle\chi\rangle e^{(d)}_e e^{(d)}_f\Big)\,.\label{ext-2pt}
\ee
Then we can immediately obtain the two-point functions by reading off the components of (\ref{ext-2pt}), namely
\bea
&& \langle T_{ij}T_{kl}\rangle=\fft{2\Theta_{ij}\Theta_{kl}}{p^4}\fft{\langle T\rangle^2}{C}
\,,\qquad \langle\chi\chi\rangle=\fft{2}{C}\langle\chi\rangle^2\,,
\cr &&
\cr && \langle T_{ij}\chi\rangle=\langle\chi T_{ij}\rangle=-\fft{2\Theta_{ij}}{p^2}\fft{\langle T\rangle
\langle\chi\rangle}{C}\,.
\eea
Provided with (\ref{T-chi-rela}) and the value of $C$ in (\ref{C}), we reproduce the results that were given by (\ref{2-pt-new}) and (\ref{ab-solu}). Obviously, (\ref{const-1}) and (\ref{const-2}) are satisfied.

\subsection{Explicit results of the two-point functions}
\label{explicittwop}

We now present the explicit two-point functions for the trace/scalar mode and these are $\langle T_{ij}T_{kl}\rangle$, $\langle T_{ij} \chi\rangle =
\langle \chi T_{ij}\rangle$ and $\langle \chi\chi\rangle$.  It should be understood that the combination
(\ref{Nij}) is a null operator and non-physical.

\subsubsection{$\langle TT\rangle$}
In momentum space, we have
\be
\langle T_{ij}(p)T_{kl}(0)\rangle=-\ft{{\rm i} (d-2) \gamma^2 \chi_s^4 \ell^{d-3}}{2^{d}\,\Gamma(\fft{d}{2})^2(d-1)(\ell^2-\gamma\chi_s^2)}
\Big(\ft{(d-2) (\ell^2+3\gamma \chi_s^2)}{d (\ell^2 + \gamma \chi_s^2)}\Big)^{\ft d2 -1} \Theta_{ij}\Theta_{kl}p^{d-4}\,,
\ee
for $d$ is odd, and
\be
\langle T_{ij}(p)T_{kl}(0)\rangle=\ft{(d-2) \gamma^2 \chi_s^4 \ell^{d-3}}{2^{d-1}\pi\,\Gamma(\fft{d}{2})^2(d-1)(\ell^2-\gamma\chi_s^2)}
\Big(\ft{(d-2) (\ell^2+3\gamma \chi_s^2)}{d (\ell^2 + \gamma \chi_s^2)}\Big)^{\ft d2 -1} \Theta_{ij}\Theta_{kl}\,p^{d-4}\log p\,,
\ee
for $d$ is even.

In the configuration space, two-point function is given by
\be
\langle T_{ij}(x)T_{kl}(0)\rangle=\ft{\Gamma(d-2)\gamma^2\chi_s^4\ell^{d-3}}{8(-1)^{\fft{d}{2}}\pi^{\fft{d}{2}+1}
\Gamma(\ft{d}{2})(d-1)(\ell^2-\gamma\chi_s^2)}\Big(
\ft{(d-2) (\ell^2+3\gamma \chi_s^2)}{d (\ell^2 + \gamma \chi_s^2)}\Big)^{\fft{d}2 -1}\hat{\Theta}_{ij}
\hat{\Theta}_{kl}\big(\fft{1}{x^{2(d-2)}}\big)\,,
\ee
where $\hat{\Theta}_{ij}$ is given by
\be
\hat{\Theta}_{ij}=\partial_i \partial_j-\eta_{ij}\Box\,.
\ee
We can obtain the explicit structure of the two-point function of energy-momentum tensor associated with the trace/scalar mode by evaluating $\hat{\Theta}_{ij}\hat{\Theta}_{kl}\big(\fft{1}{x^{2(d-2)}}\big)$, we then obtain
\bea
\hat{\Theta}_{ij}\hat{\Theta}_{kl}\big(\fft{1}{x^{2(d-2)}}\big)&=&4(d-1)(d-2)\Big(d(I_{ik}I_{jl}+I_{il}I_{jk})+
d(d-1)I_{ij}I_{kl}
\cr && -(d-1)(\eta_{ik}\eta_{jl}+\eta_{il}\eta_{jk})+4(d-3)\eta_{ij}\eta_{kl}\Big)\fft{1}{x^{2d}}\,,
\eea
where
\be
I_{ij}(x)=\eta_{ij}-\fft{2x_{i}x_{j}}{x^2}\,.
\ee

\subsubsection{$\langle\chi\chi\rangle$}

In momentum space, the two-point function is given by
\be
\langle\chi\chi\rangle=-\fft{{\rm i} (d-2)(d-1) \gamma^2 \chi_s^2 \ell^{d-3}}{2^{d+2}\,\Gamma(\fft{d}{2})^2(\ell^2-\gamma\chi_s^2)}
\Big(\fft{(d-2) (\ell^2+3\gamma \chi_s^2)}{d (\ell^2 + \gamma \chi_s^2)}\Big)^{\ft d2 -1}p^d\,,
\ee
for $d$ is odd, and
\be
\langle\chi\chi\rangle=\fft{ (d-2)(d-1) \gamma^2 \chi_s^2 \ell^{d-3}}{2^{d+1}\pi\,\Gamma(\fft{d}{2})^2(\ell^2-\gamma\chi_s^2)}
\Big(\fft{(d-2) (\ell^2+3\gamma \chi_s^2)}{d (\ell^2 + \gamma \chi_s^2)}\Big)^{\ft d2 -1}p^d\log p\,,
\ee
for $d$ is even. In configuration space, we have immediately
\be
\langle\chi(x)\chi(0)\rangle=-\fft{(d-2)(d-1)\Gamma(d+1)}{8(-1)^{\fft{d}{2}}\pi^{\fft{d}{2}+1}
\Gamma(\fft{d}{2})}\fft{ \gamma^2 \chi_s^2 \ell^{d-3}}{(\ell^2-\gamma\chi_s^2)}
\Big(\fft{(d-2) (\ell^2+3\gamma \chi_s^2)}{d (\ell^2 + \gamma \chi_s^2)}\Big)^{\ft d2 -1}\fft{1}{x^{2d}}\,.
\ee

\subsubsection{$\langle T\chi\rangle=\langle\chi T\rangle$}

In momentum space, explicitly, we have
\be
\langle T_{ij}(p)\chi(0)\rangle=-\fft{{\rm i} (d-2) \gamma^2 \chi_s^3 \ell^{d-3}}{2^{d+1}\,\Gamma(\fft{d}{2})^2(\ell^2-\gamma\chi_s^2)}
\Big(\fft{(d-2) (\ell^2+3\gamma \chi_s^2)}{d (\ell^2 + \gamma \chi_s^2)}\Big)^{\ft d2 -1} \Theta_{ij}\,p^{d-2}\,,
\ee
for $d$ is odd, and
\be
\langle T_{ij}(p)\chi(0)\rangle=\fft{(d-2) \gamma^2 \chi_s^3 \ell^{d-3}}{2^{d}\pi\,\Gamma(\fft{d}{2})^2(\ell^2-\gamma\chi_s^2)}
\Big(\fft{(d-2) (\ell^2+3\gamma \chi_s^2)}{d (\ell^2 + \gamma \chi_s^2)}\Big)^{\ft d2 -1} \Theta_{ij}\,p^{d-2}\log p\,,
\ee
for $d$ is even. In configuration space, two-point function is
\be
\langle T_{ij}(x)\chi(0)\rangle=\fft{\Gamma(d-1)(d-2) \gamma^2 \chi_s^3 \ell^{d-3}}{8(-1)^{\fft{d}{2}}\pi^{\fft{d}{2}+1}\Gamma(\fft{d}{2})(\ell^2-\gamma\chi_s^2)}
\Big(\fft{(d-2) (\ell^2+3\gamma \chi_s^2)}{d (\ell^2 + \gamma \chi_s^2)}\Big)^{\ft d2 -1} \hat{\Theta}_{ij}
\big(\fft{1}{x^{2(d-1)}}\big)\,.
\ee
It is worth noting that
\be
\hat{\Theta}_{ij}\big(\fft{1}{x^{2(d-1)}}\big)=-2(d-1)(d I_{ij}+\eta_{ij})\fft{1}{x^{2d}}\,.
\ee

To conclude this section, we stress that all the two-point functions respect the covariance of the scaling symmetry, indicating the theory is scale invariant.

\section{Additional comments in $d=2$}
\label{d=2}

As was discussed in the introduction, an scale invariant theory in $d=2$ should be enhanced to be fully conformal, satisfying the theorem proved in \cite{Polchinski:1987dy}. For the $D=3$ Einstein-Horndeski theory, we expect that the holographic two-point functions in $d=2$ for the trace/scalar modes vanish identically. Indeed, these two-point functions presented in (\ref{explicittwop}) all have an overall factor $(d-2)$.  The traceless energy-momentum tensor on the other hand has non-vanishing two-point functions and they were presented in section \ref{conformal2pt}. The overall coefficient, when specialized in $d=2$, is given by
\be
{\cal C}_T=\ell\big(1 -\fft{\gamma \chi_s^2}{\ell^2}\big)=a\,,\label{dto2}
\ee
which is precisely the holographic central charge.  Although this result fits the expectation, it is derived from holographic dictionary between the bulk graviton and boundary stress tensor in general dimensions.  However, there is no local graviton in $D=3$ gravity.

In order to derive the two-point function, we follow the procedure presented in \cite{Kraus:2018xrn}. It is convenient to introduce the holomorphic/anti-holomorphic coordinates in the AdS$_3$ boundary
\be
ds^2=\fft{\ell^2}{r^2}dr^2-r^2 dzd\bar{z}\,,
\ee
where $(z,\bar{z})=(t+x,t-x)$. We then impose the perturbation
\bea
&& ds^2=\fft{\ell^2}{r^2}dr^2-r^2 dzd\bar{z}+r^2(h_{zz}(r,z,\bar{z})dz^2+h_{\bar{z}\bar{z}}(r,z,\bar{z})d\bar{z}^2+h_{z\bar{z}}(r,z,\bar{z})dzd\bar{z})\,,
\cr && \chi(r,z,\bar{z})=-2\chi_s\log r+\psi(r,z,\bar{z})\,.
\eea
The equation of motion requires that $h_{zz}$, $h_{\bar{z}\bar{z}}$, $h_{z\bar{z}}$ and $\psi$ must take the form as
\bea
&& h_{zz}=h^{(0)}_{zz}+\fft{h^{(2)}_{zz}}{r^2}\,,\qquad h_{\bar{z}\bar{z}}=h^{(0)}_{\bar{z}\bar{z}}+\fft{h^{(2)}_{\bar{z}\bar{z}}}{r^2}\,,
\cr && h_{z\bar{z}}=h^{(0)}_{z\bar{z}}+\fft{h^{(2)}_{z\bar{z}}}{r^2}\,,\qquad \psi=\psi^{(0)}+\fft{\psi^{(2)}}{r^2}\,.
\eea
To verify that the trace part of two-point functions is indeed vanishing, we set $h_{zz}=h_{\bar{z}\bar{z}}=0$. The remaining equations of motion give rise to
\bea
&&h^{(2)}_{z\bar{z}}=-\fft{2\gamma\chi_s}{\ell^2+\gamma\chi_s^2}\psi^{(2)}\,,\nn\\ &&4\gamma^2\chi_s^3\psi^{(2)}-\ell^2(\ell^2+\gamma\chi_s^2)(
(\ell^2+\gamma\chi_s^2)\partial_z\partial_{\bar{z}}h^{(0)}_{z\bar{z}}+2\gamma\chi_s\partial_z\partial_{\bar{z}}\psi^{(0)})=0\,.
\eea
The resulting one-point functions are given by
\be
\langle T_{z\bar{z}}\rangle=-\fft{\gamma\chi_s}{4\pi\ell}\partial_z\partial_{\bar{z}}\psi^{(0)}\,,\qquad
\langle \psi\rangle=-\fft{\ell^2-\gamma\chi_s^2}{\ell\chi_s}\partial_z\partial_{\bar{z}}h_{z\bar{z}}^{(0)}\,.
\ee
These two-point functions are all vanishing up to the contact terms, i.e.
\bea
&& \fft{\delta\partial_z\partial_{\bar{z}} \psi^{(0)}}{\delta h^{(0)}_{z\bar{z}}}\sim0\,,\qquad \fft{\delta\partial_z\partial_{\bar{z}} \psi^{(0)}}{\delta \psi^{(0)}}\sim\Box\delta^2(z)\,,
\cr && \fft{\delta\partial_z\partial_{\bar{z}} h^{(0)}_{z\bar{z}}}{\delta \psi^{(0)}}\sim0\,,\qquad \fft{\delta\partial_z\partial_{\bar{z}} h^{(0)}_{z\bar{z}}}{h^{(0)}_{z\bar{z}}}\sim\Box\delta^2(z)\,.
\eea

To compute the traceless part of two-point functions, for example, $\langle T_{zz}T_{zz}\rangle$, we set $\psi=0$ and turn off the irrelevant sources $h^{(0)}_{zz}$, $h^{(0)}_{z\bar{z}}$. The remaining equations of motion reduce to
\bea
&& \partial_{\bar{z}}h^{(2)}_{zz}=\fft{\ell^2(\ell^2+\gamma\chi_s^2)}{2(\ell^2+3\gamma\chi_s^2)}\partial_z^3h^{(0)}_{\bar{z}\bar{z}}\,,
\qquad h^{(2)}_{\bar{z}\bar{z}}=\fft{\ell^2(\ell^2+\gamma\chi_s^2)}{2(\ell^2+3\gamma\chi_s^2)}\partial_{z}\partial_{\bar{z}}h^{(0)}_{\bar{z}\bar{z}}\,,
\cr &&
h^{(2)}_{z\bar{z}}=\fft{\ell^2(\ell^2+\gamma\chi_s^2)}{\ell^2+3\gamma\chi_s^2}\partial_{z}^2h^{(0)}_{\bar{z}\bar{z}}\,.
\eea
From the first equation above, we have immediately
\be
h^{(2)}_{zz}=\fft{3\ell^2(\ell^2+\gamma\chi_s^2)}{2\pi(\ell^2+3\gamma\chi_s^2)}\int d^2z' \fft{1}{(z-z')^4}h^{(0)}_{\bar{z}\bar{z}}(z',\bar{z}')\,.
\ee
(Here we made use of the formula $\partial_{\bar{z}}\fft{1}{z}=2\pi\delta^2(z)$.)
In addition, we find that the one-point function is
\be
\langle T_{zz}\rangle=\fft{\ell^2+\gamma\chi_s^2}{4\pi\ell^3}h^{(2)}_{zz}\,.
\ee
Therefore, we have
\be
\langle T_{zz}(z)T_{zz}(z')\rangle=\fft{2\delta\langle T_{zz}\rangle}{\delta h^{(0)}_{\bar{z}\bar{z}}}=\fft{3\tilde{\mathcal{C}}_T}{4\pi^2}\fft{1}{(z-z')^4}\,,
\ee
where the numerical-stripped coefficient is
\be
\tilde{\mathcal{C}}_T=\fft{(\ell^2+\gamma\chi_s^2)}{(\ell^2+3\gamma\chi_s^2)}
(1+\fft{\gamma\chi_s^2}{\ell^2})\ell\,.
\ee
This result is different from (\ref{dto2}) which was obtained from specializing the general results to $d=2$. Instead we have
\be
\tilde{\mathcal{C}}_T\simeq a+\mathcal{O}((\gamma\chi_s^2)^2)\,.
\ee
The discrepancy at the higher orders of $\gamma$ requires further investigation.

\section{Conclusion}
\label{conclusion}

In this paper, we obtained the holographic two-point functions of Einstein-Horndeski gravity with negative cosmological constant. Einstein-Horndeski gravity admits the AdS vacuum with full AdS conformal symmetry, and it is denoted as the conformal vacuum in this paper. In addition, the theory admits a scale invariant AdS vacuum whose full conformal symmetry is broken by the Horndeski scalar which exhibits the $\log r$ behavior.  Therefore, the theory should have some SQFT dual and can naturally serve as the holographic model to investigate the difference between SQFT and CFT.

We obtained the holographic two-point functions of the energy-momentum tensor in the conformal vacuum, and they are the same as those in pure Einstein-AdS gravity. Our focus was on the scale invariant vacuum. We found that the perturbations around the scale invariant vacuum have nontrivial trace/scalar mode in addition to the graviton modes. The solution is the Lorentz covariantization of the boundary cosmological mode and it can contribute to the two-point functions. We obtained the holographic counterterms associated with the scale invariant vacuum, and then we derived the two-point functions of energy-momentum tensor associated with both the graviton modes and the trace/scalar mode. The non-vanishing of the two-point function of the trace/scalar mode is a distinguishing feature of SQFTs from CFTs.

The situation becomes more subtle in $D=3$, $d=2$ case.  As expected the two-point function of the trace/scalar mode vanishes in $d=2$, indicating that the scale invariant theory is fully conformal. The central charge derived from the holographic two-point function of the energy-momentum tensor differs from holographic anomalous $a$-charge beyond the linear order of the Horndeski coupling constant $\gamma$. This discrepancy clearly warrants further investigation.

Our investigation of the scale invariant AdS vacuum in Einstein-Horndeski gravity, which is ghost free, indicates that strongly coupled scale invariant quantum field theory might exist without the full conformal invariance.  Furthermore, the operator that is dual to the trace/scalar bulk mode however violates the dominant energy condition.  Its bulk origin as the cosmological mode suggests that the boundary scalar operator may serves as an inflaton in cosmology. However, multiple subtleties remain that raise further questions. In our holographic construction, the boundary field theory must include a scalar operator that is the holographic dual of the Horndeski axion. It is this fundamental scalar operator that serves the purpose of violating the special conformal symmetry, via the trace equation (\ref{trace}). To make the bulk theory quantum complete by introducing additional necessary fields will not alter this fact unless the scale-invariant AdS vacuum no longer exists at all at the full quantum level.  On the other hand, in our construction, there is no apparent local virial current operator that is typically arising in an SQFT, indicating that the scale symmetry may be violated as well.  However, we find that the holographic two-point functions all respect the covariance of the scaling symmetry. It is thus intriguing to speculate whether there should be a generalized theory of Einstein-Horndeski gravity in which the local virial current operator is visible in the spectrum. Indeed it should be pointed that the constant shift symmetry of the Horndeski theory can be gauged to give rise to Einstein-vector theories \cite{Geng:2015kvs}, where a vector field and the curvature tensor are non-minimally coupled. It is of great interest to investigate the same issue in these theories where there are also scale-invariant but not conformal AdS vacua.

\section*{Acknolwedgement}

We are grateful to Hong-Da L\"u and Zhao-Long Wang for useful discussions. H.-Y.Z.~is grateful to the Center of Joint Quantum Studies for hospitality. This work is supported in part by NSFC grants No.~11875200 and No.~11475024.

\appendix

\section{An example in diagonal graviton modes}
\label{graviton-ex}
To illustrate the algebraic proposal in subsection \ref{subsec:algebraic}, we consider the diagonal graviton modes as an example since they are not all linearly independent, but satisfying the traceless condition. For simplicity, we focus on pure Einstein gravity in $D=4$, $d=3$, and consider following the following diagonal transverse perturbation
\be
ds^2=\fft{\ell^2}{r^2}dr^2+r^2 \eta_{ij}dx^idx^j+r^2\big(h_1(r,t) dx_1^2+h_2(r,t) dx_2^2\big)\,.
\ee
The solution is given by
\be
h_1(r,t)=-h_2(r,t)=r^{-\fft{3}{2}}H_{\fft{3}{2}}(\fft{\ell E}{r})e^{-{\rm i}Et}\,.\label{solu-Ein}
\ee
In other words, $h_1$ and $h_2$ are linearly dependent. The one-point functions can be found in \cite{Li:2018drw} and are given by
\be
\langle T_1\rangle=\fft{3}{16\pi\ell}h_1^{(d)}\,,\qquad \langle T_2\rangle=\fft{3}{16\pi\ell}h_2^{(d)}=-\langle T_1\rangle\,.\label{1-pt-Ein}
\ee
From the solution (\ref{solu-Ein}), it is clear that there is only one mode rather than two, which is similar to our case in trace/scalar mode of Einstein-Horndeski gravity at the critical point. Note one-point functions (\ref{1-pt-Ein}) are evaluated with applying the solution (\ref{solu-Ein}), hence two contributions coming from $h_1$ and $h_2$ are mixed in (\ref{1-pt-Ein}). Therefore, to compute the two-point functions, it is necessary to clarify the distinction of two parts in one-point functions. We rewrite the one-point functions as
\be
\langle T_1\rangle=a_1 h_1^{(0)}+a_2 h^{(0)}_2\,,\qquad \langle T_2\rangle=b_1 h_1^{(0)}+b_2 h^{(0)}_2\,,\label{1-pt-Ein-new}
\ee
where $a_2=b_1$ is due to $\langle T_1 T_2\rangle=\langle T_2 T_1\rangle$. We can now immediately read out the two-point functions
\be
\langle T_1 T_1\rangle=2a_1\,,\qquad \langle T_1 T_2\rangle=\langle T_2 T_1\rangle=2a_2\,,\qquad
\langle T_2 T_2\rangle=2b_2\,.
\ee
Match (\ref{1-pt-Ein}) and (\ref{1-pt-Ein-new}), giving rise to the constraints
\be
a_1-a_2=\fft{3}{16\pi\ell}\fft{h_1^{(d)}}{h_1^{(0)}}=\fft{N_1}{2}\ell^2E^3\,,\qquad b_2-a_2=\fft{3}{16\pi\ell}\fft{h_2^{(d)}}{h_2^{(0)}}=\fft{N_1}{2}\ell^2E^3\,,\label{const-1-Ein}
\ee
where $N_1$ is $\fft{{\rm i}}{8\pi}$, which is exactly the same as $N_1$ in \cite{Li:2018drw} provided with $d=3$. Moreover, (\ref{1-pt-Ein-new}) can form a matrix
\be
\left(
\begin{array}{cc}
 2a_1 & 2a_2 \\
 2a_2 & 2b_2 \\
\end{array}
\right)
\ee
The uniqueness of the source requires that the matrix should admit zero eigenvalue, then we should have
\be
a_1b_2-a_2^2=0\,.\label{const-2-Ein}
\ee
From (\ref{const-1-Ein}) and (\ref{const-2-Ein}), $a_i$ and $b_i$ can be solved
\be
a_1=b_2=\ft14{N_1}\ell^2E^3\,,\qquad a_2=b_1=-\ft14{N_1}\ell^2E^3\,.
\ee
Therefore, the two-point functions are given by
\bea
&& \langle T_1(E) T_1(0)\rangle=\ft12{N_1}\ell^2E^3\,,\qquad
\langle T_2(E) T_2(0)\rangle=\ft12{N_1}\ell^2E^3\,,
\cr &&\langle T_1(E) T_2(0)\rangle=\langle T_2(E) T_1(0)\rangle=-\ft12{N_1}\ell^2E^3\,,\label{2-pt-Ein}
\eea
On the other hand, the two-point function of Einstein gravity in momentum space is explicitly obtained, and it is given by
\be
\langle T_{ij}(p)T_{kl}(0)\rangle=N_1 \ell^2\, \fft{\Delta^{d=3}_{ijkl}(p)}{p}\,,\qquad \Delta^d_{ijkl}(p)=\ft{1}{2}(\Theta_{ik}\Theta_{jl}+\Theta_{il}\Theta_{jk}-\Theta_{ij}\Theta_{kl})\,.
\ee
We focus on $p=(E,0,0)$, and hence
\bea
&& \Theta_{11}=\Theta_{22}=E^2\,,\qquad \Theta_{12}=\Theta_{21}=0\,,
\cr && \Delta^{d=3}_{1111}=\Delta^{d=3}_{2222}=\ft{1}{2}E^4\,,\qquad \Delta^{d=3}_{1122}=\Delta^{d=3}_{2211}=-\ft{1}{2}E^4\,.
\eea
It is now clear that our results (\ref{2-pt-Ein}) match the exact results obtained in \cite{Li:2018drw}.

\end{document}